\documentclass[12pt]{article}        

\usepackage{amsmath}
\usepackage{amssymb}

\usepackage{epsfig}

\usepackage{cite}

\usepackage{epic}
\usepackage{fancybox}

\def\Order#1{${\cal O}(#1$)}

\begin{document}

\begin{titlepage}


\begin{flushright}
{\bf  
UTHEP-99-10-01}
\end{flushright}


\vspace{10mm}
\begin{center}
{\Large
Foam: Multi-dimensional General Purpose Monte Carlo Generator With Self-adapting Simplical Grid$^{\dag}$
}
\end{center}
 
\begin{center}
  {\bf S. Jadach}\\
   {\em  Department of Physics and Astronomy,\\
         The University of Tennessee, Knoxville, Tennessee 37996-1200,}\\
   {\em  DESY, Theory Group, Notkestrasse 85,  D-22603 Hamburg, Germany}\\
   { and}\\
   {\em Institute of Nuclear Physics,
        ul. Kawiory 26a, Krak\'ow, Poland}\\
\end{center}

\vspace{10mm}
\begin{abstract}
A new general purpose Monte Carlo event generator with self-adapting
grid consisting of simplices is described. 
In the process of initialization, the simplex-shaped cells
divide into daughter subcells in such a way that:
(a) cell density is biggest in areas where integrand is peaked, 
(b) cells elongate themselves along hyperspaces where integrand is enhanced/singular.
The grid is anisotropic, 
i.e. memory of the axes directions of the primary reference frame is lost.
In particular,
the algorithm is capable of dealing with distributions
featuring strong correlation among variables (like ridge along diagonal).
The presented algorithm is complementary to others known and commonly used 
in the Monte Carlo event generators.
It is, in principle, more effective than any other one for distributions with very
complicated patterns of singularities -- the price to pay is that
it is memory-hungry.
It is therefore aimed at a small number of integration dimensions ($<10$).
It should be combined with other methods for higher dimension.
The source code in Fortran77 is available from 
{http://home.cern.ch/$\sim$jadach}.
\end{abstract}

\vspace{5mm}
\begin{center}
  {\em Submitted to Comput. Phys. Commun.}
\end{center}

\vspace{10mm}
\footnoterule
\noindent
{\footnotesize
\begin{itemize}
\item[${\dag}$]
Work supported in part by Polish Government grants 
KBN 2P03B08414, 
KBN 2P03B14715, 
the US DoE contracts DE-FG05-91ER40627 and DE-AC03-76SF00515,
the Maria Sk\l{}odowska-Curie Joint Fund II PAA/DOE-97-316.
\end{itemize}
}
\begin{flushleft}
{\bf 
  UTHEP-99-10-01\\
  October 1999, revised December 1999}
\end{flushleft}

\end{titlepage}


\noindent{\bf PROGRAM SUMMARY}
\vspace{10pt}

\noindent{\sl Title of the program:} Foam, version 1.01.

\noindent{\sl Computer:}
any computer with the FORTRAN 77 compiler and the UNIX operating system

\noindent{\sl Operating system:}
UNIX, program was tested under AIX 4.x, HP-UX 10.x and Linux

\noindent{\sl Programming language used:}
FORTRAN 77 with popular extensions such as long names, etc.

\noindent{\sl High-speed storage required:}  $<$ 10 MB

\noindent{\sl No. of cards in combined program and test deck:}  2333


\noindent{\sl Nature of the physical problem:}
Monte Carlo simulation or generation of unweighted (weight equal one) events,
is a standard exercise in the particle physics, and in many other areas of the research.
It is often necessary to generate MC events according to a probability density with strong
peaks (singularities) spanned along complicated hyperspaces
of not a very well known shape.
It is highly desirable to have in the program library a general-purpose numerical tool (program)
with a MC generation algorithm featuring built-in capability of adjusting
automatically the generation procedure to an arbitrary
pattern of singularities in the probability distribution.

\noindent{\sl Method of solution:}
In the algorithm a simplical grid of vertices forming a ``foam of cells'' is built,
which adapts automatically to the integrand in such a way that the resulting ratio
of average weight to maximum weight, i.e. efficiency, is {\em arbitrarily} good.
In the subsequent MC generation the foam of cells is used to generate
one cell and a point within this cell.
The above algorithm is not based on the factorizability assumption of the integrand function
like VEGAS of P. Lepage~[1].
It can therefore generate efficiently distributions with complicated patterns of the singularities
like diagonal ridges, voids, spheres etc.

\noindent{\sl Restrictions on the complexity of the problem:}
The program is memory-hungry and therefore limited to small dimensions $n<10$.
The actual implementation is limited to $n\leq 5$, 
but only because of the function calculating the determinant.

\noindent{\sl Typical running time:}
The CPU time necessary to build up the simplical grid depends strongly on the number
of dimensions and the requested size of the grid.
On the IBM PowerPC M43P240 installation (266 MHz, 65 CERN units)
it takes about 30 secs
to build a grid of 5000 vertices,
for a simple 2-dimensional distribution.

\noindent
[1] G.~P. Lepage, J. Comput. Phys. {\bf 27},  192  (1978).

\newpage
\section{Introduction}
Generation of artificial random events within multidimensional (phase) space
according to a positive probability distribution defined by a theoretical model
is a standard exercise in the particle physics, and in many other areas of the research.
The above is usually called the ``Monte Carlo simulation'' or
generation of unweighted (weight equal one) events,
while a more modest task of calculating the integral only,
using weighted events is usually termed the ``Monte Carlo integration''.
In this work, the primary interest is in the Monte Carlo (MC) simulation,
which is a more difficult problem than the MC integration.
A computer MC program doing a MC simulation is usually called the ``MC event generator''.

With the advent of ever faster computers, one is able to
perform Monte Carlo simulation or integration in more dimensions
and for more and more complicated distributions.
All MC methods/algorithms for the efficient, i.e. fast, MC simulation/integration
can be reduced to a surprisingly small number of the basic methods
see e.g. ref.~\cite{MCguide,Jadach:1985nr}, that is to
mapping variables into more natural ones, weighting/rejecting 
and splitting the probability distribution into sum of simpler ones (branching).
For the MC event generators which are used widely, it is worth
the effort to develop a very efficient algorithm of the MC generation,
custom-made for the individual problem%
\footnote{See also ref.~\cite{James:1980} for general introduction to MC methods.}.
In this, there is no better guide than
the insight into the physics of the process to be simulated.
There are many examples of very efficient MC event generators of the custom-made type.

On the other hand, it is often necessary to perform a quick ``brute force'' MC integration
or generate MC events according to a probability density with strong
peaks (singularities) spanned along complicated hyperspaces
of not a very well known shape,
or in the case when the change of the input data induces not a very well controlled
variations in the structure of the singularities.
In all such cases it would be highly desirable to have at our disposal a numerical tool (program)
with a MC generation algorithm featuring built-in capability of adjusting
automatically the generation procedure to an arbitrary
pattern of singularities in the probability distribution.
Such a general-purpose tool was always a dream of people using MC methods.
This is an utopian dream in the sense that
we shall never get an ideal tool of this kind, i.e. working for an arbitrary distribution.
Nevertheless, we may hope to develop a MC tool with an algorithm which is fairly
efficient  for a relatively wide range of multidimensional probability  distributions.
In reality,  for each such method it is possible to find a distribution for which
the particular general-purpose MC method (tool) fails badly.
The same situation exists for the problem of finding the absolute minimum of a multi-parameter function
or in many other standard numerical problems.

The best known and widely used general-purpose algorithm for the Monte Carlo {\em integration}
of an arbitrary density function in $n$-dimensions is probably that of Lepage
described in ref.~\cite{Lepage:1978sw}, 
and embodied in the widely used Fortran code VEGAS.
In this classical algorithm, the integrand function in $n$ dimensions
is assumed to be fairly well approximated by a product of functions, 
each one depending just on one integration variable.
The  integration range of each variable is divided into $k$ bins of unequal width,
with the binning (bin sizes) different for each variable.
The entire integration domain, that is an $n$-dimensional rectangle, is divided into
$k^n$ sub-rectangles.
The whole structure is explored by means of the MC generation of random points
within each sub-rectangle, with a uniform distribution.
The result of repeated MC exploration runs is used to improve the binning.
The binning is adjusted iteratively, such that the minimum value of the ratio
of the dispersion to the average weight is achieved.
In this way VEGAS is able to do a MC {\em integration} quite efficiently.
The original VEGAS was not really aimed for the MC {\em generation}, but with a little bit of effort,
it can be adapted to the MC generation, as seen below.

As we see, the generation technique of VEGAS
is essentially an example of a multi-branching method with each branch
corresponding to one of $k^n$ rectangles.
In many practical applications (maybe even a majority) the assumption
of the factorizability of the integrand function is not violated too strongly
and the VEGAS algorithm works fine.
As expected, it fails when singularities tend to follow the diagonal of the rectangle,
i.e. variables are strongly correlated, also in the case
of big voids, singularities on the ``thin'' hyperspaces etc.
In such a cases VEGAS algorithm fails badly, and increasing the number of bins $k$,
or number of iterations, does not help to reduce weight dispersion $\sigma$ at all.
The only method to improve the integration precision is 
the brute force method of increasing the number of MC points $N$, leading to a slow
decrease of statistical error $\sim \sigma/\sqrt{N}$,
or changing analytically the integration variables (mapping).

As already stressed, our primarily interest is in the MC simulation.
The VEGAS algorithm with almost no modification can produce MC weighted events.
A little bit more effort is required to produce constant weight events,
With help of the additional rejection, knowing the maximum weight. 
This can be done by recording during the last iteration,
for each integration variable, a maximum weight in each of the $n$ bins.
The multichannel generation of the sub-rectangles
is then done using not the probability related to the average weight or its dispersion
but instead using the maximum weight (product of them).
This simple recipe works fairly well for an almost factorizable distribution.
It fails really very badly for an integrand departing from the factorizability assumption,
much worse than for the task of the MC integration only.
Essentially, the VEGAS algorithm has no means to reduce the ratio
of the maximum weight to average weight below a certain value, for a given integrand.
The only way out is then to apply mapping from the actual variables
to new ones, in which the integrand hopefully factorizes much better.
This requires detailed knowledge of the integrand distribution -- it means
going back to a labor-hungry custom-made MC.

There has been several efforts to improve on the shortcomings of the VEGAS method,
still assuming no detailed knowledge about the structure of singularities in the integrand.
For example, in ref.~\cite{Kawabata:1995th},
several improvements are done.
The most important one is adding the possibility of treating a subgroup of variables (wild)
which cause strong variation in the integrand,
while the other ones (mild) are ``averaged over''.
This is particularly useful for problems with many variables $\sim 100$,
of which  only some are ``trouble-makers'' and require special treatment.

Another improvement is described in ref.~\cite{Ohl:1999}, where
the VEGAS algorithm is upgraded with the possibility of approximating
the wild integrand not with one product of $n$ functions, each for one dimension,
but with a sum of such products, with automatic adjustments of the relative
importance of the component products.
It is essentially an application of the ideas of ref.~\cite{Kleiss:1994qy}
to the VEGAS algorithm.
The modified algorithm should be efficient for a wider class of probability distributions.

In this note, I describe an independent effort which is not rooted in VEGAS
algorithm, but rather in the algorithm used in subprogram VESKO2 of MC generator
LESKOF for deep inelastic scattering published in ref.~\cite{Jadach:1992ty}
(in fact it was already used in the much older LESKOC MC).
In VESKO2 the 2-dimensional integration area is divided into rectangular cells
which are gradually subdivided by half, along the $x$ or $y$ direction
(the choice of the division direction was random).
The division is always performed for a cell which contained the biggest
value of the integrand.
Note that this algorithm does not require factorizability of the integrand --
it is not very efficient, but numerically very much stable.
Obviously, the rule of division by half is rather primitive, one could do it better. 
The random (or arbitrary) choice of division line (along $x$ or $y$ ) could
be replaced by a better rule of dividing along the maximum gradient of the function.
However, from inspection of the way the grid of cells evolves, 
it was obvious that this algorithm has the following intrinsic problem, 
even if such improvements were implemented:
the edges of the cells are always parallel to the axes.
Consider, for instance, a narrow diagonal ``ridge'' along $x=y$ line.
Of course, the algorithm of VESKO2 is obviously superior to VEGAS, because cells
multiply and concentrate along the diagonal.
However, the adjustment of the cells would be much faster if the cells
could somehow turn around themselves to become parallel to the ``ridge''.
The self-suggesting solution is the replacement of rectangular cells with the triangular ones.
Then, hopefully in the process of subdivision, the cells could align
along singular lines, if the division rule was defined in an intelligent way.
In $n$-dimensions the generalization of triangular plaquette is simplex-shaped cell.
In the following, I shall present certain variant of such a method to which I
refer to as a ``Foam'' algorithm.

After completing the essential part of this work, I have found
in ref.~\cite{Manankova:1995xe} a description of a similar algorithm%
\footnote{ I would like to thank Viacheslav Ilyin for bringing my attention to this work.};
see last section for more comments.

The outline of the paper is the following:
Section 1 describes the Foam algorithm, 
Section 2 its implementation, 
Section 4 the usage of the program, and
Section 5 presents results of numerical tests.

\section{The Foam algorithm}

Let me define the aims which I have in mind with the new Foam algorithm:
\begin{itemize}
\item
  The algorithm is thought to be in the future a part of a bigger algorithm and it
  is supposed to take care of several ($<10$) ``wildest variables'', i.e.
  variables with the strongest singularities, while the other variables I
  imagine are dealt with the VEGAS method, or that they are ``averaged over'' like
  in BASES of ref.~\cite{Kawabata:1995th}
\item
  I assume that the integrand is completely arbitrary, in particular
  singularities may lie on arbitrarily shaped hyperspaces.
  (For extremely narrow peaks, it always make sense to map variables.)
  In particular the algorithm should be able to deal with big voids, with singularity
  along diagonals, and along ``thin'' hypersurfaces like surfaces of the cube, sphere, etc.
\item
  I imagine that in the algorithm a grid of vertices forming a ``foam of cells'' is built,
  which adapts automatically to the integrand in such a way that the resulting ratio
  of average weight to maximum weight, i.e. efficiency, is {\em arbitrarily} good.
  In the subsequent MC generation the foam of cells is used to generate
  one cell and a point within this cell.
\item
  For strong peaks the foam of cells may develop into a wrong direction,
  not knowing at the early stage of the development the positions of the sharp
  peaks containing most of the integral. I therefore require
  that the algorithm has a built-in capacity to ``collapse'' (recess) i.e.
  possibility of removing a part of the foam 
  (returning to a coarser granularity in some region).
  The iterative succession of ``grow'' and ``collapse'' should be available as an option,
  in order to stabilize the final optimal foam of cells.
\item
  The integrand should be positive and integrable. 
  Weak integrable singularities
  of the type $x^{-1/4}$ or  $\ln(1/x)$ are allowed.
  Such singularities are typically on the edges of the integration domain --
  so there should be an option to include or not the vertices at the corners
  of the simplex cell in the evaluation of the integral over the cell.
\end{itemize}

\subsection{Data structures}
The basic data structure is the {\em foam} being a linked list of {\em cells}.
A simplex cell is defined by its {\em vertices}.
Each cell  has also many other  attributes such as
pointers to parent and daughter cells, its volume, 
an estimate of the integral over the cell which I call {\em the proper integral},
average weight, maximum weight, etc.
Cells actually contain only pointers to vertices, 
while $n$-component vectors defining vertices are in a separate list of all vertices.
This organization is well justified, because one vertex may enter into several cells.
The foam is in fact a hierarchical list of cells organized into one big tree.
There are two kinds of cells, {\em inactive cells} which underwent the division
and got split into daughter cells and {\em active cells} (with no daughter).
Active cells actually cover the entire integration area.
In the MC simulation an active cell is chosen randomly according to its
{\em crude integral} which is usually bigger than its proper integral.
For the relation between crude and proper integral see below.
Each inactive cell knows the sum of the crude integrals
of all active cells it contains (all its daughter and granddaughter cells).
The aim is to make the generation of the active cell as natural
and simple as possible.
In fact, generation starts with inactive {\em root} cell at the top of the tree --
one of the daughter cells is chosen randomly according to its crude integral.
This process continues down the tree until an active cell 
(with no daughters) is randomly chosen.
The root cell is the entire integration region, being a cube of unit size.
In present algorithm, the root cell is the only one which has more than 2 daughters.
It splits into $n!$ simplices%
\footnote{This already shows why we are limited to $n<10$.
  N.B. I do not favour the other possible solution in which the unit cube is mapped
  into single simplex, because such a transformation is ``singular'' at certain vertices.}.

\subsection{Initialization: growth and collapse of the foam}
The foam structure described in previous section is constructed during the {\em initialization} phase.
It consists of subsequent {\em growths} and {\em collapses} of the foam.
Let me first describe the phase of the growth.
The initial cube is divided into  $n!$ equal simplices, daughter cells,
and each daughter cell is immediately subjected to a {\em MC exploration} procedure, e.g.
certain limited number of the trial MC events is generated within the cell
in order to calculate the average weight, dispersion, maximum weight,
minimum weight, proper integral (MC estimator) and more.
In the rest of the {\em growth} phase each cell has a chance to get divided
into 2 daughters.
In the present version of the program two options of choosing a cell for the division are implemented:
In the first method, an active cell picked up for the division is always the one which actually contains 
the biggest amount of the crude integral.
In the second method, the choice of an active cell for the division is done randomly, 
with the probability proportional to its crude integral.
The user may check empirically which option fits better his integrand.
This division process continues until the memory buffer reserved for the foam fills up%
\footnote{
  In fact, the  user may define a maximum number of cells in the foam to be smaller
  than the total length of the entire buffer. For the moment, there is no dynamical
  memory allocation in the program.}.
The active cell chosen for division, is tagged as inactive and divided into 2 daughter cells (active)
and each daughter cell undergoes the process of the MC exploration.
The recipe for the cell division is the most important part
of the algorithm; see below for its detailed description.
The division of a cell into two daughter cells involves creating a new vertex.
The new vertex is added to the list of vertices.
The sum of crude integrals calculated for the new two daughters is not necessarily equal
to the crude of the parent --
in order to maintain our algorithm of picking randomly the active cell
by descending the tree from its top to the very end of one of the branches.
The crude integral of the divided cell and of all parent cells is corrected up to the root cell, 
in such a way  that the crude integral of parent is always equal to sum of crude integrals of the daughters. 
In particular the root cell contains always the sum of the crude integrals in all active cells, 
i.e. the total crude integral, at every stage of the growth.

As already indicated, in the case of a strongly peaked distribution,
the growth may go into a ``wrong area'', 
so one is interested in a possibility of trimming/downsizing
the foam, which is termed the {\em collapse} of the foam.
The algorithm of the collapse is very simple and intuitively understandable.
When growth  is stopped by the buffer limits, the maximum value
$I^C_{\max}$ of the crude integral in all active cells is determined. 
Next, {\em all inactive} cells are checked, starting from the top cell, looking
for cells which have crude integral smaller than $I^C_{\max}$ 
times some adjustable factor close to one (the default is one, but user may reset it easily).
Every such an inactive cell is reset as the {\em active} and all its
daughters, down to the bottom of the tree, are tagged for removal.
Finally, the removal of all tagged cells is done, releasing free space in the buffer.
All vertices are also checked, to see  if they are members of any cell, and the orphan vertices
are also removed.
In this way, the entire ``un-successful'' branches are eliminated from the tree of cells,
or, in other words, several cells which are the product of the division get
replaced by the single (parent) cell, just like in the real foam!
Typically, about half of the cells are eliminated in this way, and one may start
another phase of the growth.
Note that after reviving an inactive cell, one needs to attribute to it
the original (uncorrected) crude integral.
This original crude integral from first exploration is memorized as one of attributes
of the cell, and is therefore available.
In series of the numerical tests I have found out that the collapse and subsequent growth 
usually leads to the same or very similar foam.
The above option is useful only for very sharply peaked distributions.
It is switched on only on the explicit request of the user.

\subsection{Division of the cell}
Each newly created cell undergoes exploration, just after its creation,
in order to determine its proper crude integral and the other weight parameters.
The division of the simplical cell is the essence of the algorithm.
Let me therefore describe it in a more detail.
The division procedure is defined in a maximally simple way.
A simplex of $n+1$ vertices $x_1,x_2,...,x_n,x_{n+1}$
has $n(n+1)/2$ edge lines joining every possible pair of vertices of a given cell.
In our division algorithm, one such edge between $x_i$ and $x_j$ is chosen
and the new vertex $Y$ is put somewhere on the line in between
\begin{displaymath}
 Y=\lambda x_i+(1-\lambda)x_j,\quad  0 \leq \lambda \leq 1.
\end{displaymath}
The two daughter simplices are defined with the two new list of vertices:
\begin{displaymath}
\begin{split}
& (x_1,x_2,...,x_{i-1},Y,x_{i+1},...,x_{j-1},x_j,x_{j+1},...,x_n,x_{n+1}),\\
& (x_1,x_2,...,x_{i-1},x_i,x_{i+1},...,x_{j-1},Y,x_{j+1},...,x_n,x_{n+1}).
\end{split}
\end{displaymath}
At this stage, it has do be determined which pair $(i,j)$ and which value of $\lambda$ to choose.
The aim is generally to make this choice in such a way that the function {\em varies the most strongly} 
in the direction of the edge defined by the $(i,j)$ pair of vertices.
How to find it out?
To this end, the information from the relatively
short sample of the MC events (100-1000) generated inside the cell,
during its MC exploration is exploited.
First of all, from a geometrical considerations which I omit, one is able
to ``project'' each MC point $X$ into a point $Y$ on a given edge $(i,j), i\neq j$:
\begin{displaymath}
 Y=\lambda_{ij} x_i+(1-\lambda_{ij})x_j,
\end{displaymath}
where
\begin{displaymath}
\begin{split}
&  \lambda_{ij}(X) = \frac{ |{\rm Det}_i| }{ |{\rm Det}_i| +|{\rm Det}_j| },\\
&  {\rm Det}_i = {\rm Det}(r_1,...,r_{i-1},r_{i+1},...r_n,r_{n+1}),\\
&  {\rm Det}_j = {\rm Det}(r_1,...,r_{j-1},r_{j+1},...r_n,r_{n+1}),\\
&  r_k = x_k-X,
\end{split}
\end{displaymath}
and ${\rm Det}(x_1,x_2,...,x_n)$ is the standard determinant.
The condition $0 \leq \lambda_{i,j}(X) \leq 1$ is obviously fulfilled.
With the help of the MC series of vectors $X$ (from MC exploration of the cell)
we determine for each edge $(i,j)$
the MC distribution of the variable $<\lambda_{i,j}>$,
the average $<\lambda_{i,j}>$, its variance $\sigma(\lambda_{i,j})$ etc.
For the division procedure I am looking for an edge $(i,j)$ along which
the (projected) integrand is varying most rapidly.
How do I  quantify the the ``rapidness'' of the distribution of $\lambda_{i,j}$
within the interval (0,1)?
For instance, I could use the ratio of the dispersion to the average $\sigma/<w>$ of $\lambda_{i,j}$.
This would work, if the distribution of the $\lambda_{i,j}$ had a single maximum, 
somewhere in the middle of the (0,1) interval, or at one of its ends $\lambda_{i,j}=0,1$.
This criterium of the ``rapidness'' of the distribution of $\lambda_{i,j}$
would fail, however,  if the distribution of $\lambda_{i,j}$ 
had two or more maxima within (0,1) interval.
It would be an annoying failure in many practical cases like a double ridge or closed
hyperspaces (like sphere).
A more sophisticated measure of  the ``rapidness'' of the distribution of $\lambda_{i,j}$
is therefore used in the algorithm.
For each $(i,j)$ the entire distribution $dN/d\lambda$ is recorded (histogrammed)
and the value of the integral
\begin{displaymath}
  R_{i,j} = \int \left| {dN\over d\lambda_{i,j}} - N \right| d\lambda_{i,j}
\end{displaymath}
is calculated.
The edge $(i,j)$ with the biggest value on the $R_{i,j}$ is chosen for the cell division.
As easily seen, the $R_{i,j}$ is close to zero for
flat (uniform) $dN/d\lambda_{i,j}$ and has a high value $\sim N$,
if  $dN/d\lambda_{i,j}$ features one or multiple narrow peaks.
For the division we take the value $\lambda=<\lambda_{i,j}>$.
In this way attempt is made to divide the cell into two  daughters
cells containing roughly half of the parent integral each (as in VESKO2).
In the MC exploration of the new cell, the index of the (optimal) edge $(i,j)$ 
and its $<\lambda_{i,j}>$ are readily determined and memorized for the future use --
such that when later on, this particular cell is picked up for the division,
the division direction $(i,j)$ and its ratio $<\lambda_{i,j}>$ is already predetermined.

Let me finally comment on  the weight normalization, 
and the related question of the reduction  of the variance and/or maximum weight.
In the initialization phase, the basic weight is defined
as $w= f(x) V_{Cart} $ where $f$ is integrand function,
and  $V_{Cart}$ is Cartesian volume of the cell.
The above weight is therefore normalized such
that the  proper integral is equal  the average weight, i.e.,
for infinite number of MC events $<w>$ is just equal
to the integral over the cell.

Before I enter details, it is very important to remember that my final aim
and highest priority is to generate events with the weight equal one, i.e., unweighted events.
This is a much harder task than to generate the variable weight events.
As usual, one may produce variable weight events and turn them into unweighted events
by means of rejection.
However, one cannot do it efficiently if one does not control 
very strictly the maximum weight for weighted events.
Weighted events (without strict control of the maximum weight) are only good enough for evaluating
the value of the integral.
In this less interesting case, the appropriate choice of the foam of cells, 
should also be able to minimize the variance of the weight.
Our primary aim is, however, to construct the foam of cells which will allow us to 
{\em control the maximum of the weight}, 
while decent variance of the weight is of the secondary importance.

In order to gain a good control over the maximum of the weight (and/or its variance)
I introduce the {\em crude integral} of the cell, which 
is typically an overestimated integral over the cell.
In the subsequent MC generation, the MC weight $w_{MC}$ will compensate for the fact
that the {\em crude integral} is not equal the true value of the integral over the cell.
Since the control of the maximum weight is our main priority,
in the {\em default} case, the crude integral of the cell is chosen as
\begin{displaymath}
  I_{crude} = w_{\max}=V_{Cart} {\rm Max}_{X} f(X),
\end{displaymath}
i.e., it is equal the maximum value of the integrand function times the volume of the cell.
Of course, the true maximum of the integrand function is not know,  and instead,
one takes its estimate obtained in the course of the MC exploration of the cell.
This choice ensures that the condition $w_{MC} \leq 1$, essential
for the turning the weighted events into the $w_{MC}=1$ events by means of the rejection,
will be not violated too often.

Note that if one is interested only in the variable weight events, 
for instance for calculating the integral, then a more economical
choice of the crude integral of the cell is the traditional choice
\begin{displaymath}
 I_{crude} = \sqrt{ <w>^2 + \sigma^2 } = \sqrt{<w^2>},
\end{displaymath}
i.e., the quantity which is close to the average weight $<w>$ or its variance $\sigma$,
depending which one is bigger, 
see also discussion in refs.~\cite{Lepage:1978sw,Kleiss:1994qy}.
This choice would provide a reasonable  reduction of the variance by populating more densely
cells which have bigger ratio of the proper variance to proper average weight $\sigma(w)/<w>$,
and therefore reducing the overall $\sigma(w)/<w>$.
The above choice of the crude integral is also optionally available in the program.

In either case, the compensating weight for the MC generation is always the same:
\begin{displaymath}
  w_{MC}=f(x) V_{Cart}/I_{crude}.
\end{displaymath}

\section{Program structure}
The program consists of one source file and one header file.
It is written in Fortran77 with the popular extensions like long variable names, long source lines, etc.,
which are available on all platforms.
In the {\em makefile} there is a collections of f77 compilation flags, for  Linux, AIX,
HPUX and ALPHA compilers which should be used to activate these extensions.
The program is written in such a way that its translation to c++ or Java should
be not too difficult.
In fact the program has structure of the c++ class as much as it is possible
to do it within f77.
Below I characterize the rules according to which program was written.
Variables obey the following rules:
\begin{itemize}
\item
  There is only one common block {\tt /c\_{}FoamA/} which contains all class member variables,
  which is placed in the header file {\tt FoamA.h }. 
  Each subroutine in {\tt FoamA.f } source file includes an {\tt INCLUDE 'FoamA.h' } statement.
  The outside programs should never include directly {\tt /c\_{}FoamA/ }.
  All input/output communication is done with the help of dedicated, easy to use,  subroutines.
\item
  Variables in {\tt /c\_{}FoamA/} are  {\em class members} and all have special prefix
  ``{\tt m\_{}}'' in their name, for example {\tt m\_{}Iterat} is number of iterations.
\item
  User has access to some class members through ``getters'' and ``setters''; see below.
\item
  Strong typing is imposed with help of {\tt IMPLICIT NONE}.
\end{itemize}

Subprograms in the class are loosely organized in several categories:
\begin{itemize}
\item
  Constructor with name {\tt FoamA\_{}PreInitialize} which pre-sets default values
  of many variables, including input variables.  It is invoked automatically.
\item
  Initializator with name {\tt FoamA\_{}Initialize}, which
  performs initialization of the foam grid.
\item
  Finalizator with name {\tt FoamA\_{}Finalize}, which summarizes the whole run, sets output
  values in {\tt /c\_{}FoamA/}, prints output, etc.
\item
  Maker with the name {\tt FoamA\_{}MakeSomething} or a similar one, which does the essential 
  part of job, in our case a maker {\tt FoamA\_{}MakeEvent} generates
  single MC event.
\item
  Setter with the name {\tt FoamA\_{}SetVariable}, 
  is called from the outside world to set {\tt m\_{}Variable} in {\tt /c\_{}FoamA/}.
  For example {\tt  CALL FoamA\_{}SetIterat( 5) } sets variable {\tt m\_{}Iterat=5}.
  Only certain privileged variables have a right to be served by their own setter, the other
  ones are in principle ``private''.
  Most of setters should be called before initialization.
\item
  Getter with the name {\tt FoamA\_{}SetVariable}, is called from the outside world to get 
  {\tt m\_{}Variable} from {\tt /c\_{}FoamA/}.
  It is a preferred way of sending output information to outside world.
  For example, with {\tt  CALL FoamA\_{}GetMCwt(MCwt)} one gets MC weight {\tt MCwt} in the
  user program.
\end{itemize}
The full list of class member variables in {\tt /c\_{}FoamA/}
is shown in the Appendix A.
An additional information on all subprograms of the Foam package can be found 
in Tables~\ref{tab:subprograms} and \ref{tab:subprograms2},
where I list all subprograms with short descriptions of their role.

\begin{table}
\centering
\begin{small}
\begin{tabular}{|l|p{9.0cm}|}
\hline
Subprogram & Description  \\ 
\hline\hline
\multicolumn{ 2}{|c|}{ Initialization of the foam grid}\\ \hline
FoamA\_{}PreInitialize          &Pre-initialization, set all default values (constructor)     \\
FoamA\_{}Initialize(FunW)       &Initialization of the grid, etc.                             \\
FoamA\_{}InitVertices           &Initializes first vertices of the basic cube                 \\
FoamA\_{}InitCells              &Initializes  first n-factorial cells inside original cube    \\
FoamA\_{}DefCell                &Creates new (daughter) cell and append at end of the buffer  \\
FoamA\_{}SetVertex(iVe,k1,k2,k3)&Helps to define vertex                                       \\
FoamA\_{}Explore(iCell,funW)    &Short MC sampling in iCell, determines $<wt>$, $wt_{\max}$, etc.\\
FoamA\_{}RanDiscr(Crud,nTot,Pow,iRnd) & Random choice of cell division direction              \\
FoamA\_{}MakeLambda(Lambda)     &Auxiliary procedure for FoamA\_{}Explore                     \\
FoamA\_{}Determinant(R,Det)     &Determinant of matrix R                                      \\
FoamA\_{}Det2Lapl(R,i1,i2)        & Laplace formula for 2-dim. determinant                    \\
FoamA\_{}Det3Lapl(R,i1,i2,i3)     & Laplace formula for 3-dim. determinant                    \\
FoamA\_{}Det4Lapl(R,i1,i2,i3,i4)  & Laplace formula for 4-dim. determinant                    \\
FoamA\_{}Det5Lapl(R,i1,i2,i3,i4,i5)&Laplace formula for 5-dim. determinant                    \\
FoamA\_{}Grow(funW)             &Grow cells until buffer is full                              \\
FoamA\_{}PeekMax(iCell)           & Choose randomly one cell, used also in MC generation      \\
FoamA\_{}Peek(iCell)            &Generates randomly pointer iCell of (active) cell            \\
FoamA\_{}Divide(iCell,funW,RC)  &Divide iCell into two daughters; iCell tagged as inactive    \\
FoamA\_{}Collapse               &Finds and removes some cells, revives some nonactive cells   \\
\hline \multicolumn{ 2}{|c|}{ Generation}\\ \hline
FoamA\_{}MakeEvent(funW)        &Generates point/vector Xrand with the weight MCwt            \\
FoamA\_{}GetMCvector(MCvector)   &Provides point/vector MCvector generated by MakeEvent       \\
FoamA\_{}GetMCwt(MCwt)           &Provides MC weight MCwt calculated by MakeEvent             \\
FoamA\_{}MCgenerate(funW,X,MCwt) & Alternative entry, Generates point X with the weight MCwt  \\
\hline \multicolumn{ 2}{|c|}{ Finalization}\\ \hline
FoamA\_{}Finalize(MCresult,MCerror)    &Calculates integral and its error after MC run        \\
FoamA\_{}GetIntegral(MCresult,MCerror) &Integral estimate from MC generation                  \\
\hline
\end{tabular}
\end{small}
\caption{\sf
  List of all subprograms with the short description.}
\label{tab:subprograms}
\end{table}

\begin{table}
\centering
\begin{small}
\begin{tabular}{|l|p{10.0cm}|}
\hline
Subprogram & Description  \\ 
\hline\hline
\multicolumn{ 2}{|c|}{  Other Getters and Setters}\\ \hline
FoamA\_{}GetCrude(Crude)         &Provides Crude used in MC generation                               \\
FoamA\_{}SetNdim(Ndim)           &Sets Ndim= no. of dimensions  (called before Initialize)           \\
FoamA\_{}GetNdim(Ndim)           &Provides Ndim, miscellaneous, for tests                            \\
FoamA\_{}SetnBuf(nBuf)           &Sets nBuf, length of working area in the buffer                    \\
FoamA\_{}SetIterat(Iterat)       &Sets Iterat= no. of iterations (called before Initialize)          \\
FoamA\_{}SetOut(Out)             &Sets output unit number                                            \\
FoamA\_{}SetChat(Chat)           &Sets chat level Chat=0,1,2 in the output, Chat=1 normal            \\
FoamA\_{}SetnSampl(nSampl)       &Sets nSampl; No of MC sampling before dividing cell                \\
FoamA\_{}SetOptCrude(OptCrude)   &Sets OptCrude; type of Crude =0,1,2.                               \\
FoamA\_{}SetOptBeta(OptBeta)     &Sets type of method in cell division                               \\
FoamA\_{}SetOptPeek              &Sets type of method in cell division                               \\
FoamA\_{}SetOptEdge(OptEdge)     &Sets OptEdge; (inclusion of vertices in the cell exploration)      \\
FoamA\_{}SetKillFac(KillFac)     &Sets KillFac; threshold factor for collapse procedure              \\
\hline \multicolumn{ 2}{|c|}{    Debugging and miscellaneous}\\ \hline
FoamA\_{}Check(mout,level)       &Checks all pointers (after compression) debugging!                 \\
FoamA\_{}ActUpda                 &Miscellaneous, Creates list of active cells (pointers)             \\
FoamA\_{}BufPrint(mout)          &Prints all cells, debugging                                        \\
FoamA\_{}BufActPrint(mout)       &Prints all active cells, debugging                                 \\
FoamA\_{}VertPrint(mout)         &Prints all vertices,  debugging                                    \\
FoamA\_{}PltBegin                &Plotting 2-dim. cells and vertices                                 \\
FoamA\_{}PltVert(mout)           &Plotting 2-dim. cells and vertices                                 \\
FoamA\_{}PltCell(mout)           &Plotting 2-dim. cells and vertices                                 \\
FoamA\_{}PltEnd                  &Plotting 2-dim. cells and vertices                                 \\
\hline
\end{tabular}
\end{small}
\caption{\sf
  List of all subprograms with the short description, continuation.}
\label{tab:subprograms2}
\end{table}

\begin{table}
\centering
\begin{small}
\begin{tabular}{|l|p{12.0cm}|}
\hline
Parameter & Meaning  \\ 
\hline\hline
m\_{}nDim   &  Number of dimensions.\\
m\_{}nBuf   &  Actual dynamic length of the buffer m\_{}nBuf$<$m\_{}nBufMax.
               Larger m\_{}nBuf has to be used for higher dimensions
               and for strongly singular integrand. For larger m\_{}nBuf
               the CPU time of the initialisation will increase but the total
               CPU time of the event generation will be shorter because
               the acceptance rate $<w>/w_{\max}$ will improve.
               Default is m\_{}nBuf=1000.\\
m\_{}nSampl  & Number of MC sampling per cell in the MC {\em exploration} of the new cell daughter
               cell. The MC efficiency $<w>/w_{\max}$ seems to depend weakly on m\_{}nSampl.
               However, if one cannot increase m\_{}nBuf any more then enlarging m\_{}nSampl
               may still help a little bit.
               Default is m\_{}nSampl=200.\\
m\_{}Iterat &  No. of iterations in the initialization of the grid, 
               m\_{}Iterat=0 is the lowest possible value. In most cases it is enough.
               Each iteration consists of the grow and collapse of the grid.
               Several iterations are recommended for very strongly peaked distributions.
               Default is m\_{}Iterat=0.\\
m\_{}KillFac & Threshold factor for reviving inactive cells in the ``collapse'' procedure
               of the iteration. Its change seems to be without much effect.
               May be in some rare cases the user will find profitable to readjust it.
               Default is m\_{}KillFac=1.\\
m\_{}OptCrude& Type of the crude integral used for the MC generation of the active cell.
               For OptCrude=0 estimator of the ``true'' integral in the cell is taken as the crude,
               for OptCrude=1 the value of $\sqrt{<w^2>}$ and for OptCrude=2 the maximum weight
               $w_{\max}$. Default is m\_{}OptCrude=2.\\
m\_{}OptEdge & Option parameter deciding whether vertices are included in the MC exploration
               of the cell. For m\_{}OptEdge=0 they are not included
               and for m\_{}OptEdge=1 they are included. Generally it is good
               to include vertices, but if there are some weak singularities or numerical
               instabilities of the integrand close to boundary of the integration domain,
               then it is necessary to set m\_{}OptEdge=0.
               Default is m\_{}OptEdge=1.\\
m\_OptPeek   & Option parameter for the method of selecting the cell for division.
               OptPeek=0 cell with maximum crude (default), OptPeek=1 randomly.\\
m\_OptBeta  & Type of choice of the edge in the division of the cell, 
               Default OptBeta=0 described in the text, OptBeta=1, OptBeta=2 for tests.\\
m\_{}Out     & Output unit number. For redirecting output from Foam to separate disk file.
               Default is m\_{}Out=6.\\
m\_{}Chat    & Chat=0,1,2 increasing chat level in the output unit, 
               Chat=1 is the default normal level\\
\hline
\end{tabular}
\end{small}
\caption{\sf
  Important input parameters of the Foam. 
  They are listed in the order of their importance.}
\label{tab:input}
\end{table}

\newpage
\section{Program usage, input parameters}
Basic input variables are listed in
in Table~(\ref{tab:input}) together with their explanation.
Typical user program using Foam package may look as follows:
\vspace{-2mm}
{\small
\begin{verbatim}
*------------------------------------------------------------------------
  DOUBLE PRECISION   Density
  EXTERNAL           Density
  CALL FoamA_SetNdim(       3) ! number of dimensions
  CALL FoamA_SetnBuf(    2000) ! length of buffer
  CALL FoamA_SetIterat(     1) ! number of iterations
  CALL FoamA_SetnSampl(   500) ! no. of MC events/cell (initialization)
  CALL FoamA_SetOptCrude(   2) ! type of crude, =2 is default anyway
  CALL FoamA_SetOptEdge(    1) ! edge point are included, (=0 excluded)
  CALL FoamA_SetChat(       1) ! printout level
  CALL FoamA_Initialize(Density) ! initialize foam grid
  DO loop = 1, 200000
     CALL FoamA_MakeEvent(Density)     ! generate MC event
     CALL FoamA_GetMCvector(MCvector)  ! get MC event, vector
     CALL FoamA_GetMCwt(MCwt)          ! get MC weight
     CALL GLK_Fil1(1000, MCwt,1d0)     ! users histogramming
  ENDDO
  CALL FoamA_Finalize(MCresult,MCerror) ! printouts, get integral & error
  CALL FindWtLimit(1000)         ! users routine, check on MC efficiency
*------------------------------------------------------------------------
\end{verbatim}}
\vspace{-2mm}
\noindent
In fact the user has to set only the number of dimensions {\tt Ndim}.
The other input variables 
{\tt nBuf, Iterat, nSampl, OptCrude, OptEdge, Chat} are already preset for the user,
thus calling the setters for them is optional.
The user needs to provide his own integrand function, which in this example is {\tt Density}.
Below is shown an example of a simple integrand function (3-dim. sphere).
\vspace{-2mm}
{\small
\begin{verbatim}
*------------------------------------------------------------------------
      DOUBLE PRECISION FUNCTION Density(X)
*////////////////////////////////////////////////////////////////////////
*// 3-dimensional testing function, Thin sphere centred at (A1,A2,A3) //
*// with Radius and Thickness defined below                            //
*////////////////////////////////////////////////////////////////////////
      IMPLICIT NONE
      DOUBLE PRECISION  X(*)
      DOUBLE PRECISION  Radius,Thickness,A1,A2,A3,R
      DATA   A1,A2,A3  / 0.25, 0.40, 0.50 / ! centre of sphere
      DATA   Radius    / 0.35  /            ! radius  of sphere
      DATA   Thickness / 0.020 /            ! thickness of sphere
*------------------------------------------------------------------------
      R   = SQRT( (x(1)-A1)**2 +(x(2)-A2)**2 +(x(3)-A3)**2 )
      Density = Thickness/( (R-Radius)**2 + Thickness**2)
      END
*------------------------------------------------------------------------
\end{verbatim}

\section{Numerical tests}
A minimum of testing of the program and the algorithm was done.
It is even more important that the program was already tested 
in a {\em real practical application}.
The present version of Foam is implemented as a part of {\cal KK}
Monte Carlo event generator  \cite{KKcpc:1999}
for the up to 3-dimensional problem of the simulation of the initial state
photon radiation (ISR) and the beamstrahlung for the future Linear Colliders%
\footnote{
  It this practical application Foam is more efficient than VEGAS by the factor $>10$.}.

In the following I present the comparisons of the Foam
program with VEGAS~\cite{Lepage:1978sw} for $n=2,3$.
For $n=2$ I use the following three testing functions:
\begin{equation}
\begin{split}
\label{eq:test-functions}
f_a(x_1,x_2)&= 1-\Theta(0.5 -|x_1-0.5|-\gamma)\; \Theta(0.5 -|x_1-0.5-\gamma|),\; \gamma=0.05,\\
f_b(x_1,x_2)&= {1/4\pi R^2}\;
               {\gamma  \over \pi [ (R- \sqrt{(x_1-0.25)^2 +(x_2-0.40)^2})^2  +\gamma^2 )] },\;
               \gamma =0.02, R=0.35  \\
f_c(x_1,x_2)&= {\gamma \over \pi [ (x_1+x_2)^2 +\gamma^2 )]},\; \gamma =0.02.  \\
\end{split}
\end{equation}
All above functions are defined within the unit square $0\leq x_i \leq 1$.
The first density function peaks along one of the diagonals of the square,
the second one 
peaks along a 2\% wide ring centered at (0.25,0.40) of the radius 0.25,
and the last one
represents 5\% wide band all along the four edges of the unit square.

\begin{figure}[!th]
\centering
\setlength{\unitlength}{0.1mm}
\begin{picture}(1600,1600)
\put( 350,1550){\makebox(0,0)[t]{\hbox{\Large (a)}}}
\put( 400, 750){\makebox(0,0)[t]{\hbox{\Large (b)}}}
\put(1200, 750){\makebox(0,0)[t]{\hbox{\Large (c)}}}
\put(  400, 800){\makebox(0,0)[lb]{\epsfig{file=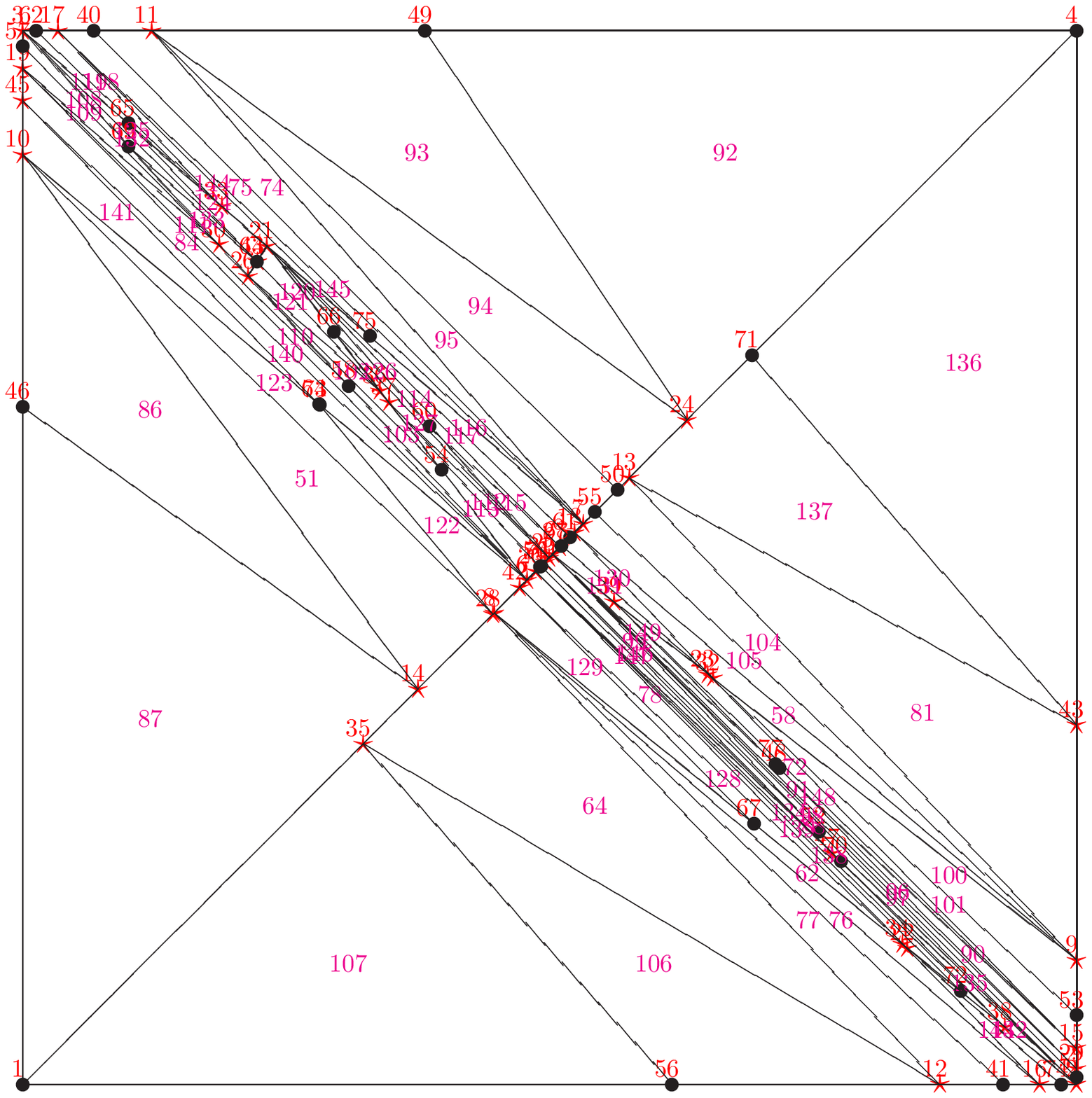,width=80mm,height=80mm}
}}
\put(  -20,   0){\makebox(0,0)[lb]{\epsfig{file=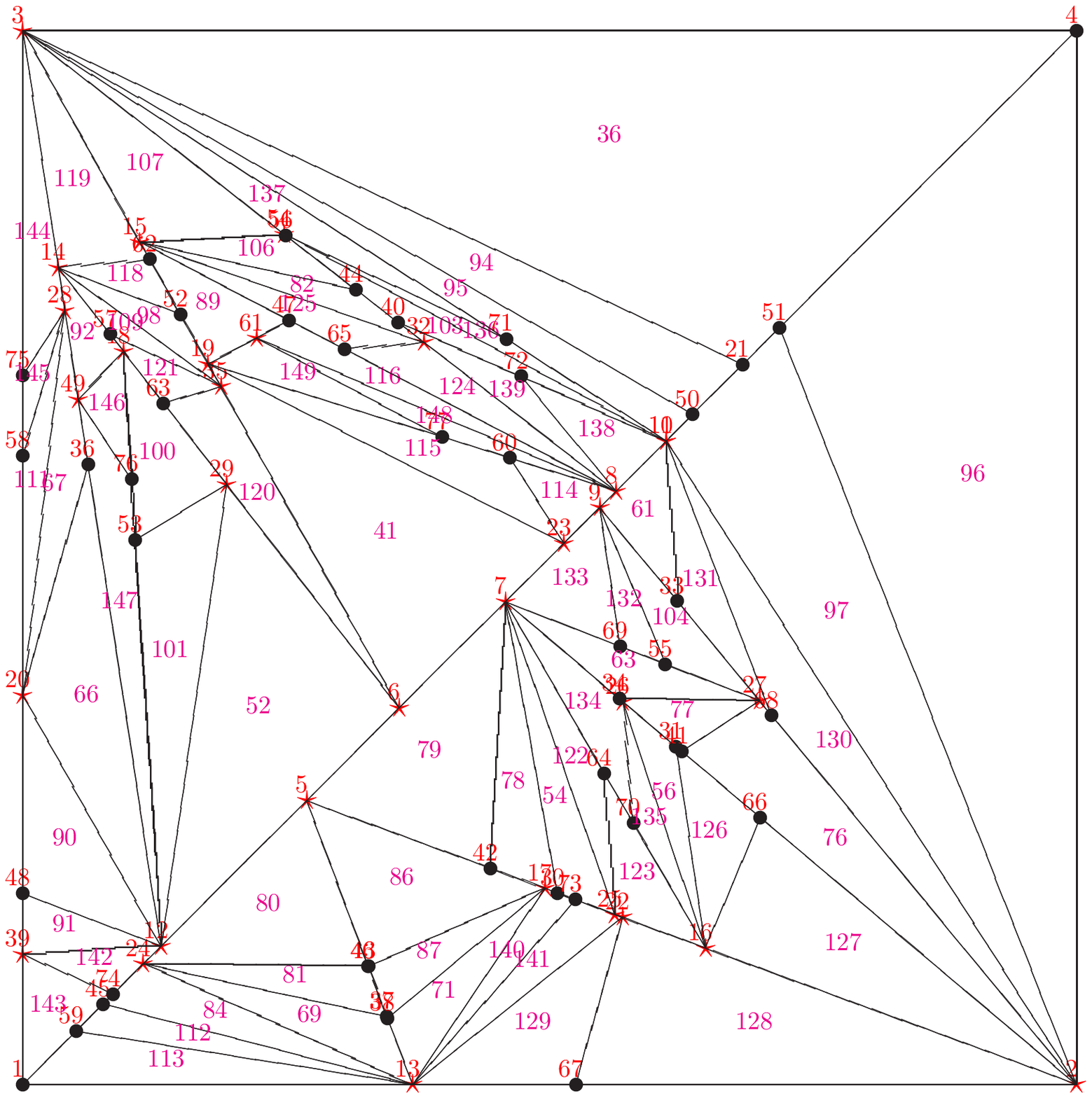,width=80mm,height=80mm}
}}
\put(  800,   0){\makebox(0,0)[lb]{\epsfig{file=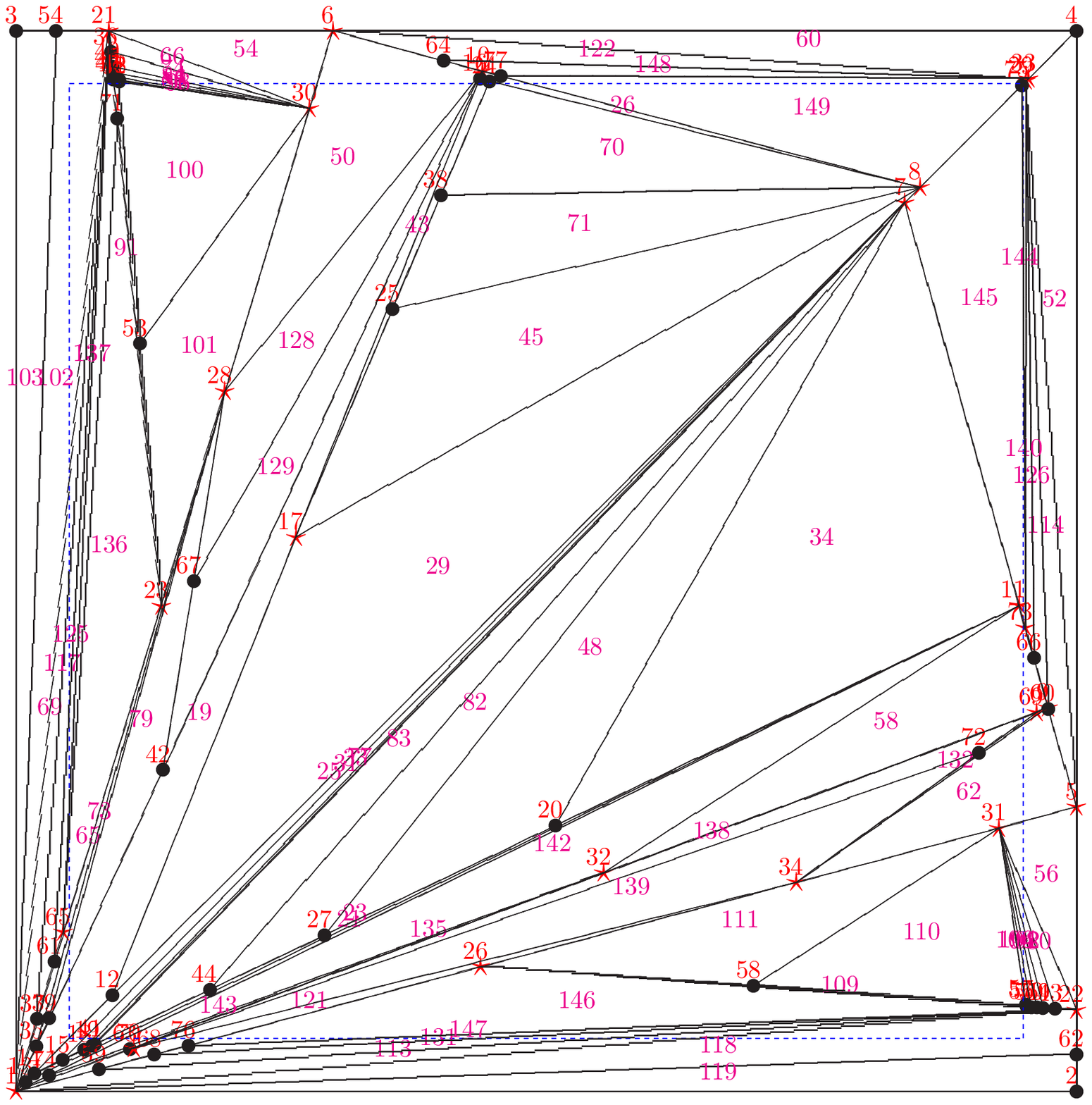,width=80mm,height=80mm}
}}
\end{picture}
\caption{\small\sf
  Two dimensional foam of the cells for three density functions $f_a$, $f_b$ and $f_c$
  defined in eq.~\protect(\ref{eq:test-functions}).
  For $f_a$ the boundary of the nonzero integrand is marked with a dashed line.
}
\label{fig:foam}
\end{figure}

In Fig.~\ref{fig:foam} the resulting 2-dimensional foam of cells is plotted. 
In each case, the foam consists of about 75 active cells and the exploration of
the single cell in the initialization was based on
the 1000 MC events per cell.
Only the active cells are plotted.
As expected, the cells of the foam concentrate in the areas of the enhancement
of the integrand functions.
They clearly try tend to {\em elongate} along the lines of
the ``ridges'' in the integrand functions -- exactly as desired and expected.
The elongation is especially well pronounced in the case (a) of the diagonal ridge.

\begin{table}[!th]
\centering
\begin{small}
\begin{tabular}{|l|r|r|}
\hline
Functions                  &  Foam    & VEGAS  \\ 
\hline\hline
$f_a(x_1,x_2)$ (diagonal ridge)     &  0.94  & 0.05 \\
$f_b(x_1,x_2)$ (circular ridge)     &  0.83  & 0.15 \\
$f_c(x_1,x_2)$ (edge of square)     &  0.57  & 0.53 \\
\hline
\end{tabular}
\end{small}
\caption{\sf
  The efficiency  $w_{\max}^\varepsilon$, for $\varepsilon=10^{-4}$,
  of Foam and VEGAS for 3 examples of the 2-dimensional integrand function
  defined in eq.~\protect(\ref{eq:test-functions}).
  After the initialization, efficiency was determined from a sample of the $10^6$ MC events.
  The results are from the Foam run with the 5000 cells (about 2500 active cells) 
  and the cell exploration was based on 200 MC events per cell.
}
\label{tab:eff2dim}
\end{table}

The three test functions  of eq.~(\ref{eq:test-functions})
are intended to be of the ``non-factorizable'' type, such that
Foam should be more efficient than VEGAS.
Let me stress that
{\em it is definitely my aim to adjust the concept of ``efficiency'' of the MC
to the task of MC generation of weight one events.}
(It should not be confused with the statistical error of the integral.)
Because of that, I define the efficiency as the ratio of the average weight
to maximum weight $<w>/w_{\max}$, 
such that it is equal to the rejection rate in the process of turning
variable-weight events into $w=1$ events.
In practice, however, $w_{\max}$ has to be defined unambiguously
and in a numerically stable way.
The straightforward definition of $w_{\max}$ as a maximum weight 
determined empirically in the MC test run,  or during the initialization of the grid,
can be prone to large fluctuation.
For practical reasons I do not want to exclude from our considerations
the case of the  weak integrable singularities in the integrand
function like, for instance,  $x^{-1/10}$ which may lead to a tail in the weight distribution.
(Numerical instabilities due to the rounding errors may produce a similar effect.)
Quite often, such a tail will not influence the average weight $<w>$ 
and the MC estimate of the integral at all.
It will, however, render $w_{\max}$ ill-defined, that is dependent on the number of the events
in the MC sample and/or wildly fluctuating.
In order to avoid such a problem, in all presented tests,
the following alternative definition of the $w_{\max}$ is applied:
For a given precision level $\varepsilon<<1$,
the $w_{\max}$ is determined from the weight distribution
in such a way, that the contribution to the $<w>$ (that is to the total integral)
from ``under-rejected'' (``over-weighted'') events with  $w>w_{\max}$ is equal $\varepsilon$.
Such a quantity is referred to%
\footnote{ 
  The concept of such a maximum weight was already used in BHLUMI MC \protect\cite{bhlumi2:1992}.}
as $w_{\max}^\varepsilon$.
In practice it is a little bit of effort to determine the $w_{\max}^\varepsilon$ from the weight distribution.
One has to create a histogram for the weight distribution
with at least $\sim 1000$ bins, in order to determine
the $w_{\max}^\varepsilon$ with at least 2-digit precision.

In order to understand correctly  the following
comparisons of the Foam and VEGAS,
it is important to remember that Foam has always certain intrisic advantage over VEGAS.
For Foam the efficiency $<w>/w_{\max}^\varepsilon$ can be improved further and further by 
means of a brute force increase of the number of cells and/or manipulating its other input parameters,
while for VEGAS it is not possible.
More precisely in practical applications in the MC simulation
one has to face the following fundamental deficiency of VEGAS:
For a given type of the integrand function, its efficiency is limited in a rigid way --
the increase of function calls and/or number of iteration in VEGAS cannot improve the efficiency
$<w>/w_{\max}^\varepsilon$  beyond certain asymptotic value which I call {\em asymptotic efficiency}.
This annoying limitation is excluding VEGAS from many practical MC applications.
In order to the improve efficiency of VEGAS beyond the above asymptotic value, 
it is necessary to map the integration variables or employ the other
MC techniques like branching etc.
However, this is exactly contrary to the spirit of a general purpose MC tool like VEGAS or Foam.

In Tab.~\ref{tab:eff2dim} the MC efficiency of Foam and VEGAS is compared for
three testing integrand functions of eq.~(\ref{eq:test-functions}).
As expected, Foam is significantly more efficient for $f_a$ and $f_b$
which clearly do not fulfil the factorizability assumption.
The case of $f_b$ is a little bit more complicated; see below.
Tab.~\ref{tab:eff2dim} shows the asymptotic efficiencies for VEGAS,
that is the best possible one. 
Even worse, for both functions $f_a$ and $f_b$ these asymptotic efficiencies of VEGAS are of \Order{\gamma}
and can be even worse  smaller $\gamma$ parameter (sharper singularities).
As indicated, the case of the $f_c$ is more involved.
Although at first sight $f_c$ looks clearly a non-factorizable, however, it is
surprisingly well approximated by the product of two functions and its asymptotic efficiency
is equal $1/2 + $\Order{\gamma}.
Even more interestingly, 
although for the Foam algorithm there is no limiting asymptotic efficiency 
(its asymptotic efficiency for large number of cells is arbitrarily close one), 
nevertheless the increase of number of cells for $f_c$ 
does lead to a rather slow improvement of its efficiency.
This phenomenon is also seen in Fig.~\ref{fig:foam}(c) where the foam of cells
fits visually the shape of $f_c$ worse than in the other two cases%
\footnote{
  Apparently, a grid with rectangular cells would do better in this case.}.
The amount of CPU time necessary for producing the results in Tab.~\ref{tab:eff2dim}
was about the same for the Foam and VEGAS programs.

\begin{table}[!th]
\centering
\begin{small}
\begin{tabular}{|l|r|r|}
\hline
Functions                          &  Foam  & VEGAS  \\ 
\hline\hline
$f_a(x_1,x_2,x_3)$ (thin diagonal)     &  0.67  & 0.04 \\
$f_b(x_1,x_2,x_3)$ (thin sphere)       &  0.36  & 0.10 \\
$f_c(x_1,x_2,x_3)$ (surface of cube)   &  0.37  & 0.33 \\
\hline
\end{tabular}
\end{small}
\caption{\sf
  The efficiency  $w_{\max}^\varepsilon$, for $\varepsilon=10^{-4}$,
  of Foam and VEGAS for 3 examples of the 3-dimensional integrand functions, the analogs of
  the integrand of eq.~\protect(\ref{eq:test-functions}).
  After the initialization, efficiency was determined from a sample of the $10^6$ MC events.
  Results from Foam are for the 5000 cells (about 2500 active cells) and the cell exploration was based
  on 200 MC events per cell.
}
\label{tab:eff3dim}
\end{table}
In Tab.~\ref{tab:eff3dim} results of comparison of Foam and VEGAS  are shown  for the 3-dimensional
integrands being straightforward extension to 3-dimensions
of the functions of eqs.~(\ref{eq:test-functions}).
Again Foam is clearly superior, 
and again the efficiency of VEGAS is already equal its limiting/asymptotic value,
while the efficiency of Foam can be still improved by adding more cells in the initialization phase.
In this 3-dimensional case the CPU time consumption of Foam is noticeably smaller than of VEGAS.

As it was already discussed in the description of the Foam algorithm,
the actual implementation has several operational modes corresponding to several
variants of the algorithm.
The user may switch to one of them by changing input parameters/switches.
All these operational modes were tested and the default configuration corresponds
to a mode which is the best, according to a presently accumulated experience.
One disappointing result of these tests of various operational modes
is that the iterative grow/collapse of the foam does not really improve the grid.
Nevertheless I leave this option in the actual program, 
because it might be a useful
as a starting point for some fresh idea in the future development of the algorithm.

Concerning similar algorithms proposed in ref.~\cite{Manankova:1995xe}
it may be noted that the algorithm presented here is defined 
more narrowly, but it is also defined very clearly, while in ref.~~\cite{Manankova:1995xe}
authors propose the whole family of algorithms exploiting the idea of simplical grid,
without going much into details.
In particular they contemplate an interesting possibility
of adapting the grid to the integrand, by moving vertices and deforming cells.
It is an attractive idea, but it is difficult to judge how difficult it would be to implement 
it and how important a gain would be in the efficiency,
without defining fine details of the algorithm.
The other idea in ref.~\cite{Manankova:1995xe} is that of division of a cell in several cells,
while I deliberately limit myself to division into only two cells.
Even the division into two cells can be realized in several ways, and
I have realized two scenarios with the resulting efficiency being quite different.
Without a clear guiding idea, entering into a game of multi-division of cells 
looks like taking risk of getting lost in hundreds of options and scenarios.
It is also important to remember that the main concern in the present work is MC simulation with
the constant weight events, while  ref.~\cite{Manankova:1995xe} is rather aiming
at the easier task of the MC integration with weighted events.
This pre-determines the priorities in the construction of the algorithm very strongly and differently.

\section{Conclusions}
A new general purpose Monte Carlo tool based on a simplical self-adapting grid
is available.
Numerical tests show that for the non-factorizable integrand functions it is
more effective than the classical VEGAS solution.
The first real application to beamstrahlung and ISR in the {\cal KK} Monte Carlo
confirms maturity of the solution.
The main limitation is that the algorithm and the program
is well adapted to a relatively small dimensions, $n<8$.
The algorithm and the program should not be treated as a final solution --
it is rather a beginning of a new development direction in the MC methods.

\vspace{15mm}
{\bf\Large Acknowledgements}

\vspace{1mm}
Most of this work was done during my visit in DESY --
I am greatly indebted to DESY Directorate for its generous support.
I am also thankful for support of the Physics Department of the University of Tennessee,
where the final version this work was prepared.
I would like to thank T. Ohl, W. P\l{}aczek, F. Tkachov and S. Yost for reading the manuscript
and useful comments.

\newpage
\section{Appendix A}
Below the entire ``class common block'' of the Foam class is show explicitly.
\vspace{-2mm}
{\small
\begin{verbatim}
      COMMON /c_FoamA/   
     $ m_CeStat(m_nBufMax),           ! Cell member: status=0 inactive, =1 active
     $ m_CePare(m_nBufMax),           ! Cell member: parent cell pointer
     $ m_CeDau1(m_nBufMax),           ! Cell member: daughter1 cell pointer
     $ m_CeDau2(m_nBufMax),           ! Cell member: daughter2 cell pointer
     $ m_CeVert(m_nBufMax,m_NdiMax+1),! Cell member: vertex pointers
     $ m_CeIntg(m_nBufMax),           ! Cell member: integral estimator
     $ m_CeCrud(m_nBufMax),           ! Cell member: Crude integral estimate
     $ m_CeVolu(m_nBufMax),           ! Cell member: Cartesian volume
     $ m_CeXave(m_nBufMax),           ! Cell member: Average best X
     $ m_CeBest(m_nBufMax),           ! Cell member: Best pair of vertices, pointer
     $ m_CeSum( m_nBufMax,m_sMax),    ! Cell member: weight summaries
     $ m_VerX(  m_vMax, m_NdiMax), ! List of all VERTEX positions
     $ m_ActC(m_cMax),             ! List of all pointers to ACTIVE cells
     $ m_VolTot,                   ! Estimate of Volume total, without error
     $ m_Crude,             ! M.C. generation Crude value of integral
     $ m_SumWt,             ! M.C. generation sum of Wt
     $ m_SumWt2,            ! M.C. generation sum of Wt**2
     $ m_NevGen,            ! M.C. generation sum of 1d0
     $ m_WtMax,             ! M.C. generation maximum wt
     $ m_WtMin,             ! M.C. generation minimum wt
     $ m_MCresult,          ! M.C. generation Final value of ITEGRAL
     $ m_MCerror,           ! M.C. generation Final walue of ERROR
     $ m_MCwt,              ! M.C. generation current event weight
     $ m_MCvector(m_NdiMax),! M.C. generated vector
     $ m_KillFac,           ! Threshold factor for collapse of cells
     $ m_Ndim,              ! dimension of the problem
     $ m_nBuf,              ! Actual dynamic lenth of the buffer m_nBuf<m_nBufMax
     $ m_LastVe,            ! Last vertex
     $ m_LastAc,            ! Last active cell
     $ m_LastCe,            ! Last cell in buffer 
     $ m_nSampl,            ! No. of sampling when dividing cell
     $ m_Iterat,            ! No. of iterations in consolidation process
     $ m_Ncalls,            ! No. of function calls, total
     $ m_OptCrude,          ! type of Crude =0,1,2 for TrueVol,Sigma,WtMax
     $ m_OptEdge,           ! decides whether vertices are included in the sampling
     $ m_Chat,              ! Chat level in output, Chat=1 normal level
     $ m_Out                ! Output unit %$
\end{verbatim}


\end{document}